\documentclass[twocolumn,showpacs,preprintnumbers,amsmath,amssymb,superscriptaddress]{revtex4-1}
\usepackage{graphicx}
\usepackage{dcolumn}
\usepackage{bm}
\usepackage{CJK, CJKnumb}
\usepackage{color}
\begin{document}
\begin{CJK*}{GBK}{song}

\title{Antiferromagnetic superlattices: anisotropic band and spin-valley valve in buckled two-dimensional materials}
\author{Wei-Tao Lu} \email{physlu@163.com}
\affiliation{School of Physics and Technology, Nantong University, Nantong 226019, China}
\author{Tie-Feng Fang}
\affiliation{School of Physics and Technology, Nantong University, Nantong 226019, China}
\author{Qing-Feng Sun} \email{sunqf@pku.edu.cn}
\affiliation{International Center for Quantum Materials, School of Physics, Peking University, Beijing 100871, China}
\affiliation{Hefei National Laboratory, Hefei 230088, China}

\begin{abstract}
Antiferromagnetic superlattices (AFSL) are proposed based on the buckled hexagonal two-dimensional materials, which can be realized by the proximity effect of the periodically deposited antiferromagnets.
It is found that the AF proximity effect can give rise to valley-polarized minibands and conductance, which are not held under ferromagnetic proximity.
The spin degeneracy and valley degeneracy are lifted simultaneously in the presence of AF proximity and electric field.
In consequence, both minibands and conductance could be spin-valley polarized completely in AFSL.
The symmetry of spin-valley polarization is analysed by considering the pseudospin rotation operations and spatial inversion operations.
Furthermore, AFSL also induce a highly anisotropic band structure due to the spin-orbit coupling (SOC).
In particular, the group velocity parallel to the periodic direction of AFSL is greatly renormalized, while the velocity perpendicular to the periodic direction remains unaffected, contrary to that observed in graphene superlattices.
With the increase of SOC, the anisotropy becomes more prominent, leading to flattened band and electron supercollimation.
The direction of anisotropy can be regulated by adjusting the potential and SOC.
These findings offer an alternative approach to engineering anisotropic two-dimensional materials.
As an application, the AFSL may well work as a symmetry-protected spin-valley valve easily controlled by the gate voltages.
\end{abstract}
\maketitle

\section{Introduction}

Recently, antiferromagnetic (AF) spintronics has attracted tremendous attention due to its potential applications for high-density and ultrafast information
devices \cite{Baltz, HYan}.
Experimental results demonstrate that an AF memory can be both written and read electrically \cite{Wadley, Han}.
Spin transport is the crucial issue of spintronics.
Many works on the transport property are reported in various AF systems, such as AF topological insulators \cite{Deng, Liang}, AF tunnel junctions \cite{Nunez, Zelezny, Lu, Qin}, antiferromagnet/superconductor junctions \cite{Jakobsen, Lu2, Chourasia}, and AF hexagonal lattices \cite{Qiao, Stepanov, Hogl, Miao, addr1, Ezawa, Marin, MLuo, XXu, XZhai}.
The quantized Hall resistance proved that quantum anomalous Hall effect could be observed in the AF topological insulator MnBi$_2$Te$_4$ \cite{Deng}.
N\'{u}\~{n}ez et al. predicted theoretically that the giant magnetoresistance and spin-transfer torque will occur in the AF tunnel junction, which may work as an AF spin valve \cite{Nunez}.

AF spintronics in two-dimensional (2D) materials is also studied widely \cite{Qiao, Stepanov, Hogl, Miao, addr1, Ezawa, Marin, MLuo, XXu, XZhai}.
By introducing AF order to one half of the silicene nanoribbon, the helical edge states are present only on the other half of the nanoribbon, leading to a perfect spin filter \cite{Ezawa}.
A tunnel-field-effect spin filter is proposed based on interband tunneling in the AF stanene \cite{Marin}.
Depending on the configuration of the conducting layer and substrates, three types of AF spin valves can be realized in 2D hexagonal lattices \cite{MLuo}.
Graphene and other hexagonal 2D materials have another degree of freedom, that is valley, and the two valleys are related by the time-reversal symmetry.
It is found that in silicene the valley could be used to mediate a spin-diode effect through the ferromagnetic/AF junction \cite{XZhai}.
In the AF silicene/superconductor junction, a spin-valley filter is realized between the crossed Andreev reflection and elastic cotunneling by adjusting the electric field \cite{Lu2}.
Besides, the spin-valley filtering effect can also be realized in strained graphene \cite{Peeters} and ferromagnetic silicene \cite{GJin} by considering  spin-orbit coupling (SOC).

Superlattice (SL) is an effective strategy for engineering the electronic structure in semiconductors and 2D materials.
The transport behavior and energy band tailored by SL are helpful for optimizing the performance of devices.
A number of SL artificially designed by periodic potentials \cite{Park2, Park, Brey, Barbier, WHKang}, local strains \cite{Lu3, YZhang}, ferromagnetic fields \cite{Missault, Missault2, CHChen, Rojas, Hajati}, and circularly polarized lights \cite{Hajati, YGuo,YGuo2} are proposed in 2D materials.
It has been predicted that the magnetic-strained SL may break the time-reversal symmetry in graphene, and so the energy band and transport strongly depend on the valley degree of freedom, which can be used to construct a valley filter device \cite{Lu3}.
The ferromagnetic SL can enhance the spin and valley polarizations and realize an electric field-controlled switching of the current in silicene \cite{Missault}.
In addition, owing to the chiral nature of graphene, the propagation of charge carriers through a SL potential is highly anisotropic, and the group velocities are reduced to zero in one direction but are unchanged in another \cite{Park2}.
The specific anisotropy and chirality of the graphene SL make it a natural candidate for electron supercollimation \cite{Park}.
Besides, extra zero-energy Dirac points are generated in graphene SL \cite{Brey, Barbier}, which can be further cloned to higher energies \cite{WHKang}.
The spin degenerate energy band and Dirac points can be broken by a ferromagnetic SL \cite{Missault2}.
In experiment, it is observed that the graphene samples exhibit high transport anisotropy between the current directions parallel and perpendicular to the SL vector, with extra Dirac points emerging in the former configuration \cite{YLi, TLi}.
Even so, there is still a lack of investigation on the antiferromagnetic superlattices (AFSL) formed by the AF exchange fields, and the AFSL induced electronic properties are expected to be studied.

In this work, we propose the AFSL in a buckled 2D materials.
For the buckled 2D hexagonal materials, such as silicene and germanene, the sublattices A and B are separated from each other in the direction perpendicular to the sheet plane.
The buckled structure allows for the band gap to be tunable by the perpendicular electric field \cite{Drummond, ZNi}.
The considered AFSL consist of a series of AF exchange fields and perpendicular electric fields, as shown in Fig. 1.
Experimentally, a local AF exchange field can be realized via the proximity effect by depositing the antiferromagnet, such as MnPSe$_3$, on the 2D material \cite{XLi, Hogl}.
The local electric field can be modulated by the top and bottom gate voltages applied on the buckled lattices \cite{Drummond, ZNi}.
The main findings of this work are as follows:
(1) The AF exchange field alone could lead to valley-polarized conductance and minibands, but they are spin-polarized in the ferromagnetic SL \cite{Missault, Missault2, CHChen, Rojas, Hajati}.
(2) The spin-valley valve effect for minibands and transport can be achieved in AFSL, which shows certain symmetry with respect to AF exchange field and electric field, providing specific signatures for AFSL.
(3) Importantly, the AFSL induce an anomalous anisotropic band structure due to SOC. The energy band for AFSL possesses a highly flattened energy dispersion in the direction parallel to the SL period, contrary to results for SL potential in graphene where the anisotropic band is flattened in the direction perpendicular to the SL period \cite{Park2, YLi}.
Because of SOC and AF exchange field, the anisotropy and Dirac point are spin-valley dependent. 

The rest of this paper is organized as follows.
In Sec. II we present the theoretical formalism and dispersion relation of the system.
The numerical results on spin-valley-polarized conductance, anisotropic energy band, and the corresponding symmetry analysis are shown in Sec. III.
A brief summary is presented in Sec. IV.

\section{Theoretical Formulation}

In the presence of AF exchange field and perpendicular electric field, the physics of a buckled 2D hexagonal material could be well described by a nearest-neighbor tight-binding model  \cite{CCLiu, Ezawa}.
At low energy, the effective Hamiltonian for electronic states near Dirac points can be written as \cite{Hogl, Ezawa, CCLiu}:
\begin{align}
H_{\eta s}=H_0 + (\lambda_z + s \lambda_{AF} - \eta s \lambda_{SO}) \sigma_z + U,
\end{align}
and $H_0=\hbar v_F (k_x \sigma_x - \eta k_y \sigma_y)$ is the Hamiltonian for pristine graphene.
$\sigma=(\sigma_x, \sigma_y, \sigma_z)$ are the Pauli matrices in the sublattice $A$ and $B$ spaces, $v_F$ the Fermi velocity,
$\eta=\pm 1$ and $s= \pm 1$ describe the valleys $K$/$K'$ and
the spin up/spin down of the electrons, respectively.
$\lambda_{SO}$ is the strength of SOC, $\lambda_{AF}$ is the AF exchange field, $U$ is a gate-controlled chemical potential, and $\lambda_z=\ell E_z$ is the staggered sublattice potential induced by perpendicular electric field $E_z$ with $2\ell$ the vertical separation of A and B sublattices \cite{Drummond, Ezawa}.
As shown in Fig. 1, we set that the widths of the AF and normal regions in AFSL are $d_1$ and $d_2$, respectively. There are no AF exchange field and electric field ($\lambda_{AF}= \lambda_z =U=0$) in the normal region.
Note that the direction of AFSL is assumed to be perpendicular to the direction between the two valleys, so there is no intervalley scattering.

\begin{figure}
\includegraphics[width=8.0cm,height=3.0cm]{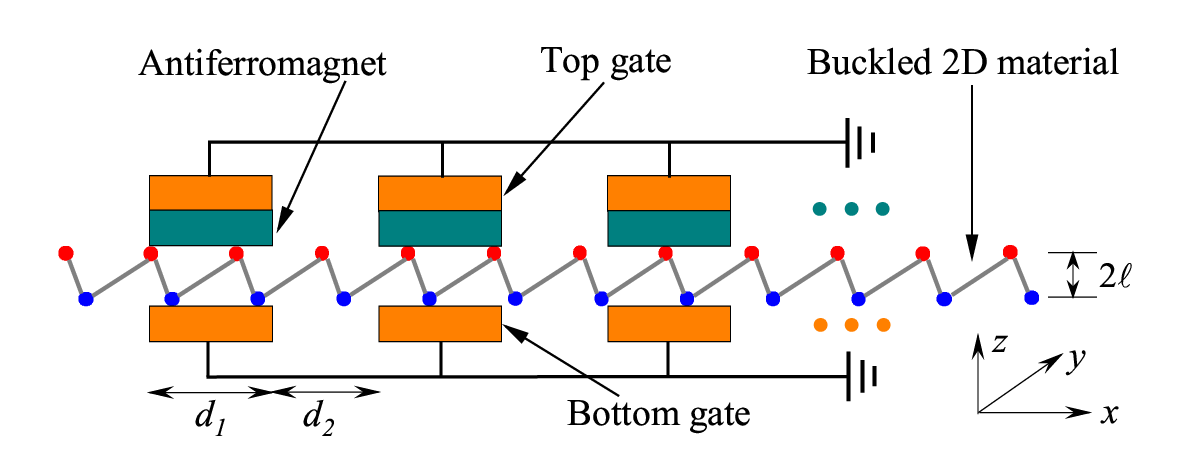}
\caption{ The schematic diagram for AFSL model on the buckled 2D materials, which consists of periodically modulated antiferromagnet and gate voltages.
AFSL have periodic variation along the $x$ direction.
The length of one unit is $d=d_1+d_2$ with barrier width $d_1$ and well width $d_2$.
The vertical distance between the sublattices A and B is $2\ell$.}
\end{figure}

The eigenvalue of Hamiltonian (1) is given by
\begin{align}
E_{\eta s} = U \pm \sqrt {\Delta_{\eta s}^2 + (\hbar v_F k_F)^2},
\end{align}
with the momentum $k_F$ and $\Delta_{\eta s}=\lambda_z + s \lambda_{AF} - \eta s \lambda_{SO}$, describing a spin and valley related band gap.
Due to the conservation of transverse wave vector $k_y$ in the one-dimensional AFSL, the two-component spinor envelope function with incident energy $E$ in the $j$th region of AFSL has the form
\begin{align}
\psi_j(x) = G_j F_j \left(\begin{array}{cc} a_j \\ b_j \end{array}\right) e^{i k_y y}
\end{align}
where $G_j=\left(\begin{array}{cc} 1  &  1  \\ k_-/\epsilon_j & -k_+/\epsilon_j \end{array}\right)$,
$F_j=\left(\begin{array}{cc} e^{i q_j x}  &  0  \\ 0 &  e^{-i q_j x} \end{array}\right)$,
and $k_\pm=\hbar v_F(q_j \pm i \eta k_y)$.
In the AF region $\epsilon_j=\epsilon_1=E-U+\Delta_{\eta s}$ and $q_j=q_1=\sqrt{[(E-U)^2-\Delta_{\eta s}^2]/(\hbar v_F)^2-k_y^2}$,
otherwise, $\epsilon_j=\epsilon_2=E - \eta s \lambda_{SO}$ and $q_j=q_2=\sqrt{(E^2-\lambda_{SO}^2)/(\hbar v_F)^2-k_y^2}$.
$q_j$ and $k_y$ are components of $k_F$ in the $x$ and $y$ directions.
$a_j$ and $b_j$ are the transmission and reflection coefficients, respectively.

Based on the continuity condition of wave functions at the interface $x=x_j$ between the $j$th and $(j+1)$th regions, $\psi_j |_{x_j}=\psi_{j+1} |_{x_j}$,
one can get: $\left(\begin{array}{cc} a_{j+1} \\ b_{j+1} \end{array}\right) = M_j \left(\begin{array}{cc} a_j \\ b_j \end{array}\right)$, and $M_j = F_{j+1}^{-1}(x_j) G_{j+1}^{-1} G_j F_j(x_j)$. Thus, the total transfer matrix for the system with $N$ regions can be written as $M=M_{N-1} \cdots M_j \cdots M_1$. Then the transmission probability can be obtained from
\begin{align}
T_{\eta s} = 1 - |M_{21}|^2/|M_{22}|^2,
\end{align}
and $M_{ij}$ is the matrix element of $M$. Using Landauer-B\"{u}ttiker formula, the spin- and valley-dependent conductance at zero temperature is given by
\begin{align}
G_{\eta s}(E)=G_0 \int_{-\pi/2}^{\pi/2} T_{\eta s}(E,E\sin \theta) \cos \theta d \theta,
\end{align}
where $\theta$ is the incident angle with respect to the $x$ direction, $G_0=2 e^2 E L_y / (\pi \hbar)$ is taken as the conductance unit, and $L_y$ is the sample size along the $y$ direction.

For the infinite AFSL, the dispersion relation can be also calculated by the transfer matrix method \cite{Lu3}.
According to Bloch's theorem, the spin- and valley-dependent dispersion relation $E_{\eta s}(k_x)$ is achieved by solving the transcendental equation,
\begin{align}
&\cos(k_x d) = \cos(q_1 d_1) \cos(q_2 d_2) - \chi  \sin(q_1 d_1) \sin(q_2 d_2), \\
&\chi = \frac{(\epsilon_1 q_2)^2 + (\epsilon_2 q_1)^2 + (\epsilon_1 - \epsilon_2)^2 k^2_y}{2 \epsilon_1 \epsilon_2 q_1 q_2}
\end{align}
with the Bloch wave number $k_x$ and the unit length of AFSL $d=d_1+d_2$.
Eqs. (6) and (7) suggest that the dispersion relation is invariant with respect to $k_x \rightarrow -k_x$ and $k_y \rightarrow -k_y$.
The spin-valley-dependent electronic spectrum is expected because of AF exchange field and electric field.

\section{Results and discussions}
We first present results for the AF exchange field induced valley-polarized conductance and minibands in subsection III A.
Then the spin-valley valve effect and the symmetry relations are studied in subsection III B.
Subsection III C gives the results for anisotropic band structure and spin-valley-dependent Dirac points.
The incident energy is set as $E=5.0meV$ for convenience in the following discussion.
In subsections III A and B, $\lambda_{SO}=4.0meV$ which is the SOC strength of silicene \cite{CCLiu}.

\subsection{Valley valve effect}

\begin{figure}
\includegraphics[width=8.0cm,height=8.0cm]{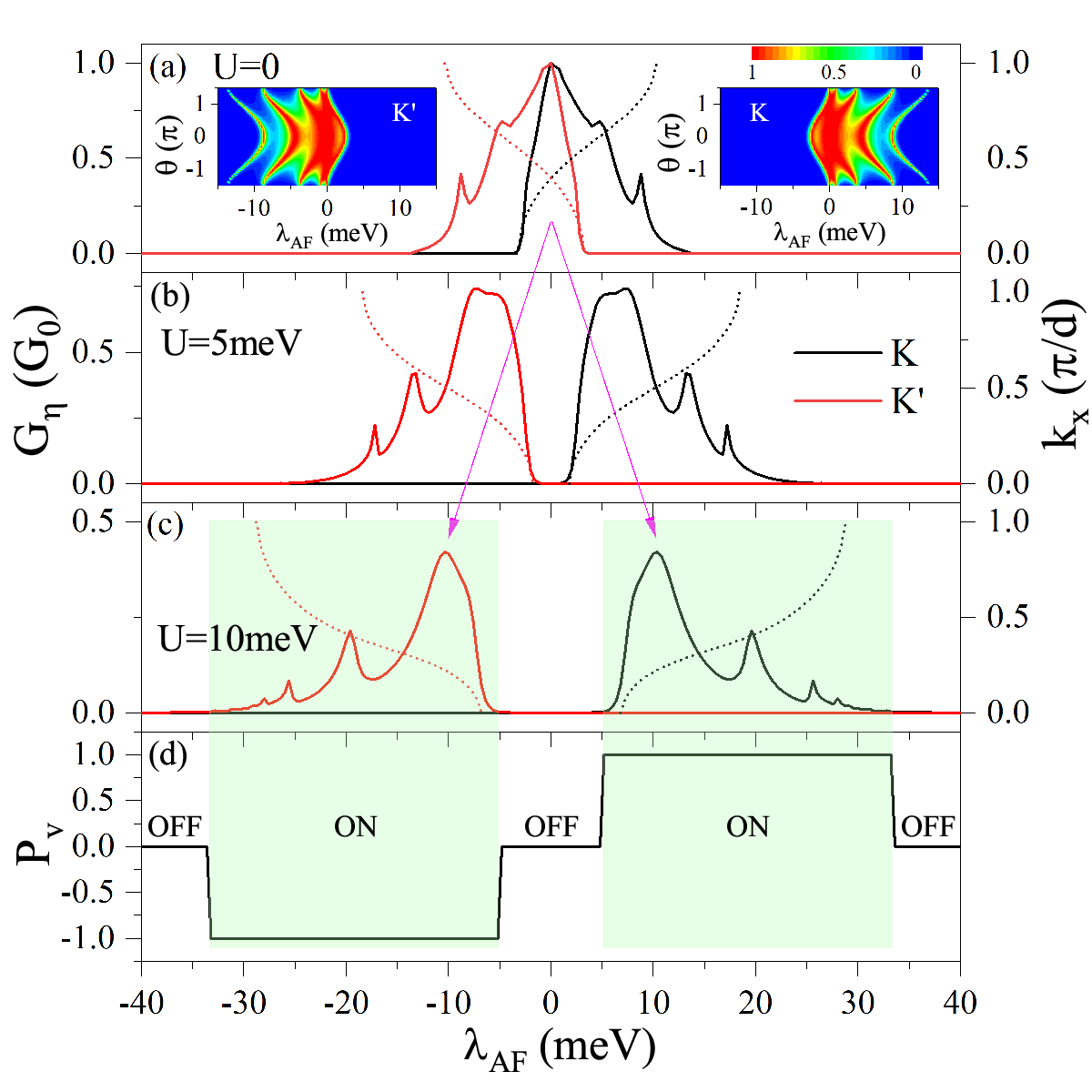}
\caption{ (a-c) Conductance $G_\eta$ (solid curves) and minibands (dashed curves) versus AF exchange field $\lambda_{AF}$ when $\lambda_z=0$, $n=5$, $d_1=50nm$, and $d_2=100nm$. $n$ is the period number of AFSL. The inset of (a) is the contour plot of corresponding transmission probability $T_\eta (\lambda_{AF}, \theta)$. (d) Valley polarization $P_v$ for the conductance in (c). }
\end{figure}

First, we discuss the minibands and conductance of AFSL in the absence of electric field with $\lambda_z=0$.
The eigenvalue is simplified as $E_\eta = U \pm \sqrt {(\lambda_{AF} - \eta \lambda_{SO})^2 + (\hbar v_F k_F)^2}$, which is dependent on valley index, but not on spin index, due to SOC.
Thus, from Fig. 2 we can see that the AF exchange field gives rise to
valley-polarized minibands and valley-polarized conductance but without spin polarization,
that is $G_K=G_{K\uparrow}=G_{K\downarrow}$, $G_{K'}=G_{K'\uparrow}=G_{K'\downarrow}$, but $G_K \neq G_{K'}$, 
which can be also understood by the isolated minibands in Fig. 7(b).
Note that the ferromagnetic field alone cannot induce the valley polarization \cite{Missault, Missault2, CHChen, Rojas, Hajati}.
Figs. 2(a)-2(c) indicate that with the increase of potential $U$, the conductance $G_K$ gradually moves to the right along the $\lambda_{AF}$ axis, while $G_{K'}$ moves to the left, and so the conductances from different valleys are completely separated.
The minibands change in the same way as the conductances.
Both minibands and conductances of the two valleys are symmetric with respect to $\lambda_{AF}=0$, i.e., $G_K(\lambda_{AF})=G_{K'}(-\lambda_{AF})$.
In addition, the conductance exhibits a resonance behavior, resulting from the resonant transmission probability $T_\eta (\lambda_{AF}, \theta)$ [see the inset of Fig. 2(a)].
$T_\eta (\lambda_{AF}, \theta)$ is symmetric about $\theta=0$.
Obviously, the conductance peak corresponds to the resonance region of $T_\eta$, which arises from the resonant modes in the quantum wells of AFSL.
Fig. 2(d) shows the valley polarization $P_v$ for the conductance in Fig. 2(c), which is defined as $P_v=(G_K-G_{K'})/(G_K+G_{K'})$.
One may find a perfect platform of $P_v$ and it is antisymmetric with respect to $\lambda_{AF}=0$.
As the AF exchange field $\lambda_{AF}$ increases, $P_v$ is $0$, $-1$, $0$, $1$, and $0$ in sequence, and the window of perfect valley polarization can be turned on and off. So AFSL can well act as a valley valve.

Symmetry is important to the experimental design and theoretical research.
The above symmetry of minibands and conductance can be analysed by considering the pseudospin rotation operators $\sigma_{x,y,z}$ and the spatial inversion operators $R_{x,y}$.
In fact, the system AFSL always have a spatial inversion symmetry related with the operator $\mathcal{M}_1=\sigma_z R_x R_y$.
The Hamiltonian satisfies $\mathcal{M}_1 H_{\eta s} \mathcal{M}_1^{-1} = H_{\eta s}$ and $\mathcal{M}_1 \psi_{\eta s} (k_x, k_y) = \psi_{\eta s} (-k_x, -k_y)$, giving rise to $T_{\eta s} (k_y) = T_{\eta s} (-k_y)$.
Thus, the transmission probability $T_{\eta s}$ is always symmetric about $k_y=0$ [see the inset of Fig. 2(a)], independent of $\lambda_z$, $\lambda_{AF}$, and $U$.
When $\lambda_z=0$ and $\lambda_{AF}\neq0$ for Fig. 2, the Hamiltonian between the two spins is invariant under transformation
\begin{align}
\mathcal{M}_2 H_{\eta s} (\lambda_{AF}) \mathcal{M}_2^{-1} = H_{\eta \bar{s}} (\lambda_{AF})
\end{align}
with $\mathcal{M}_2=\sigma_x R_y$ or $\sigma_y R_x$ and $\bar{s}=-s$. Then the conductances satisfy $G_{\eta s} (\lambda_{AF})=G_{\eta \bar{s}} (\lambda_{AF})$, i.e., $G_{K\uparrow}=G_{K\downarrow}$ and $G_{K'\uparrow}=G_{K'\downarrow}$.
For the two valleys, under the operator $\mathcal{M}_3=\sigma_x$ or $\sigma_y R_x R_y$, the Hamiltonian has
\begin{align}
\mathcal{M}_3 H_{\eta s} (\lambda_{AF}) \mathcal{M}_3^{-1} = H_{\bar{\eta} s} (-\lambda_{AF})
\end{align}
with $\bar{\eta}=-\eta$.
It can be concluded that the conductances between the two valleys are related by $G_{\eta s} (\lambda_{AF})=G_{\bar{\eta} s} (-\lambda_{AF})$, corresponding to $G_K(\lambda_{AF})=G_{K'}(-\lambda_{AF})$ and $P_v(\lambda_{AF})=-P_v(-\lambda_{AF})$ in Fig. 2.

\subsection{Spin-valley valve effect}

\begin{figure}
\includegraphics[width=8.0cm,height=3.0cm]{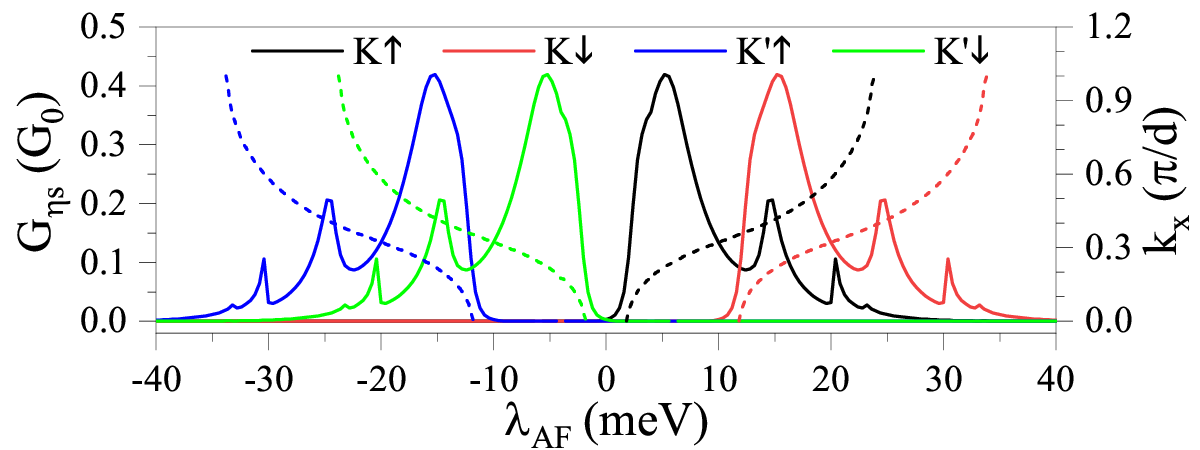}
\caption{ Conductance $G_{\eta s}$ (solid curves) and minibands (dashed curves) versus AF exchange field $\lambda_{AF}$ when $\lambda_z=5meV$. The values of other parameters are the same as these in Fig. 2(c).}
\end{figure}

The existence of electric field $\lambda_z$ would break the transformations
in Eqs. (8) and (9), and lift the spin degeneracy, leading to the conductances $G_{\eta s}(\lambda_{AF}) \neq G_{\eta \bar{s}}(\lambda_{AF})$ and $G_{\eta s} (\lambda_{AF}) \neq G_{\bar{\eta} s} (-\lambda_{AF})$.
As a result, the positions of minibands and conductance strongly depend on the spin and valley indexes, which occur in different regions of $\lambda_{AF}$, as shown in Fig. 3, 
consistent with the results in Fig. 7(c) in the following.
However, the system satisfies
\begin{align}
\mathcal{M}_4 H_{\eta s} (\lambda_{AF}) \mathcal{M}_4^{-1} = H_{\bar{\eta} \bar{s}} (-\lambda_{AF})
\end{align}
with $\mathcal{M}_4=\sigma_z R_x$, and so $G_{\eta s} (\lambda_{AF})=G_{\bar{\eta} \bar{s}} (-\lambda_{AF})$.
This means that the conductances $G_{K\uparrow}$ and $G_{K'\downarrow}$ (or $G_{K\downarrow}$ and $G_{K'\uparrow}$) are symmetric with respect to $\lambda_{AF}=0$, and so is the minibands (see Fig. 3).

\begin{figure}
\includegraphics[width=8.0cm,height=9.0cm]{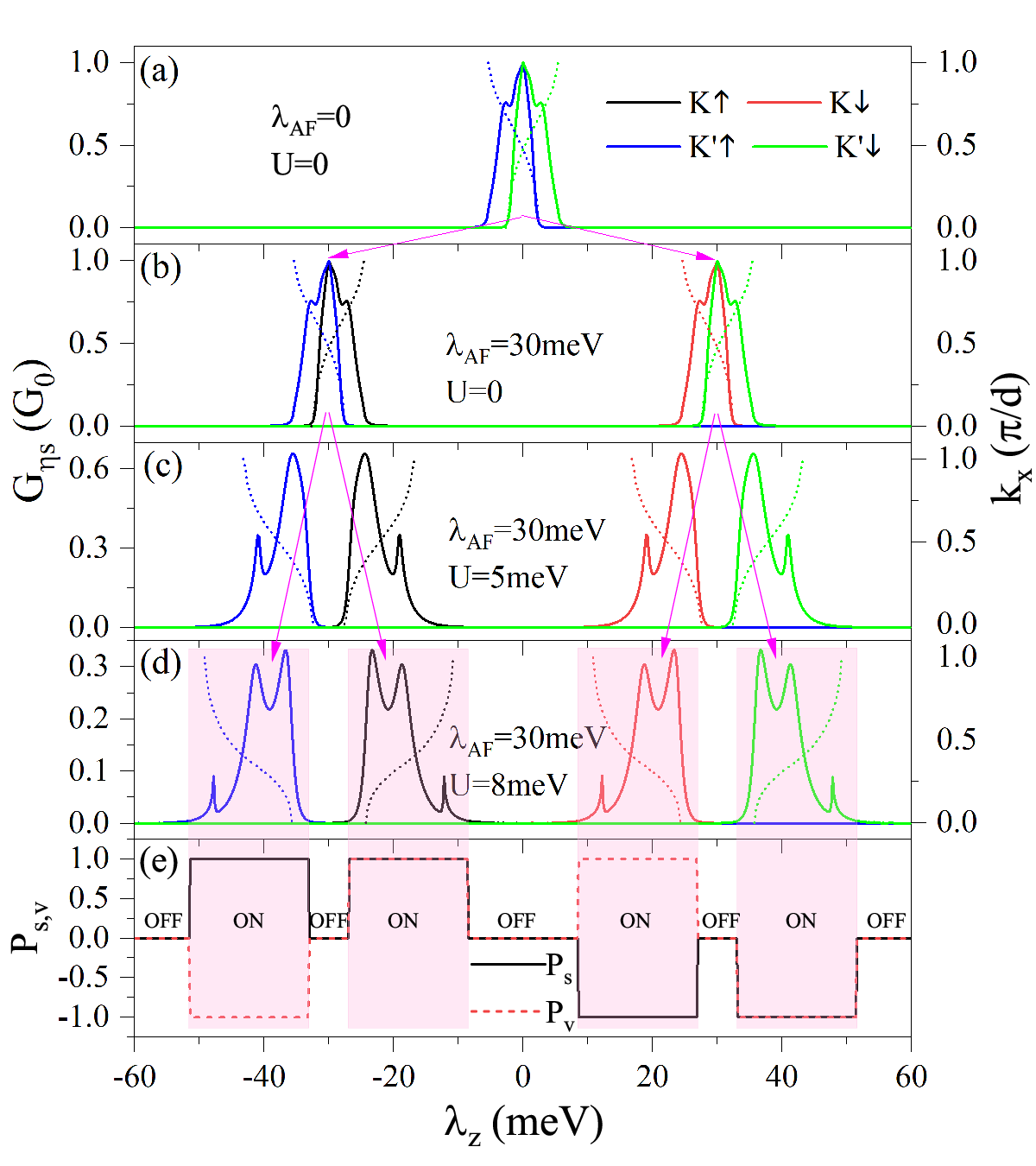}
\caption{ (a-d) Conductance $G_{\eta s}$ (solid curves) and minibands (dashed curves) versus electric field $\lambda_z$ when $n=3$, $d_1=80nm$, and $d_2=100nm$. (e) Spin and valley polarizations $P_{s,v}$ for the conductance in (d).}
\end{figure}

Fig. 4 presents the spin-valley-polarized minibands and conductance as a function of electric field $\lambda_z$.
The four kinds of conductances are the same and equal to $G_0$ at $\lambda_z=\lambda_{AF}=0$ [see Fig. 4(a)].
When $\lambda_z\neq0$ and $\lambda_{AF}=0$ in Fig. 4(a), the system has $\mathcal{M}_4 H_{\eta s} (\lambda_z) \mathcal{M}_4^{-1} = H_{\bar{\eta}\bar{s}} (\lambda_z)$ and $G_{\eta s} (\lambda_z)=G_{\bar{\eta} \bar{s}} (\lambda_z)$.
In addition, the Hamiltonian also satisfies
\begin{eqnarray}
\mathcal{M}_2 H_{\eta s} (\lambda_z) \mathcal{M}_2^{-1} &=& H_{\eta \bar{s}} (-\lambda_z),\\
\mathcal{M}_3 H_{\eta s} (\lambda_z) \mathcal{M}_3^{-1} &=& H_{\bar{\eta} s} (-\lambda_z),
\end{eqnarray}
leading to the relations of conductances $G_{\eta s} (\lambda_z) = G_{\eta \bar{s}} (-\lambda_z)$ and $G_{\eta s} (\lambda_z) = G_{\bar{\eta} s} (-\lambda_z)$, respectively.
Thus, the conductances satisfy $G_{K\uparrow}=G_{K'\downarrow}$ and $G_{K\downarrow}=G_{K'\uparrow}$, which are symmetric with respect to $\lambda_z=0$, as observed in Fig. 4(a).
Obviously, there are no spin and valley polarizations in this case.

With the appearance of AF exchange field in Figs. 4(b)-4(d), the time-reversal symmetry for spin and valley is destroyed.
The AF exchange field $\lambda_{AF}$ lifts the spin degeneracy and valley degeneracy simultaneously in the presence of SOC and electric field.
Particularly, the minibands and conductances for up spin move to $\lambda_z=-\lambda_{AF}$, while the minibands and conductances for down spin move to $\lambda_z=\lambda_{AF}$.
As the potential $U$ increases in Figs. 4(b)-4(d), the minibands and conductances for different spins and valleys are completely separated.
The minibands and conductances exhibit interesting symmetry.
The conductances for up spin $G_{K\uparrow}$ and $G_{K'\uparrow}$ are symmetric with respect to $\lambda_z=-\lambda_{AF}$.
For down spin, $G_{K\downarrow}$ and $G_{K'\downarrow}$ are symmetric about $\lambda_z=\lambda_{AF}$.
For $K$ (or $K'$) valley, $G_{K\uparrow}$ and $G_{K\downarrow}$ (or $G_{K'\uparrow}$ and $G_{K'\downarrow}$) are symmetric about $\lambda_z=0$.
That's because Eq. (11) and $G_{\eta s} (\lambda_z) = G_{\eta \bar{s}} (-\lambda_z)$ are still valid when $\lambda_{AF}\neq0$.
In the presence of AF exchange field $\lambda_{AF}$, the transformation (12) for the two valleys would become
\begin{align}
\mathcal{M}_3 H_{\eta s} (\lambda_z-s\lambda_{AF}) \mathcal{M}_3^{-1}
= H_{\bar{\eta} s} (-\lambda_z - s \lambda_{AF})
\end{align}
and one may get $G_{\eta s} (\lambda_z - s \lambda_{AF}) = G_{\bar{\eta} s} (-\lambda_z - s \lambda_{AF})$.
This suggests that the conductances between the two valleys are symmetric about $\lambda_z=-s \lambda_{AF}$ depending on the spin index.
The spin-valley-polarized minibands also have the above symmetric relationships [see Figs. 4(b)-4(d)].
In consequence, the perfect spin-valley polarization is achieved, as shown in Fig. 4(e).
Here, the spin and valley polarizations are defined as
\begin{eqnarray}
P_s&=&(G_{K\uparrow}+G_{K'\uparrow}-G_{K\downarrow}-G_{K'\downarrow})/G_{total},\\
P_v&=&(G_{K\uparrow}+G_{K\downarrow}-G_{K'\uparrow}-G_{K'\downarrow})/G_{total}
\end{eqnarray}
with $G_{total}=(G_{K\uparrow}+G_{K'\uparrow}+G_{K\downarrow}+G_{K'\downarrow})$.
Because of the operators $\mathcal{M}_2$ and $\mathcal{M}_3$,
the spin and valley polarizations obey $P_s(\lambda_z)=-P_s(-\lambda_z)$
and $P_v(\lambda_z)=P_v(-\lambda_z)$, respectively.
$P_s$ is antisymmetric about $\lambda_z=0$, while $P_v$ is symmetric about $\lambda_z=0$.
Furthermore, the symmetric relationships described by Eqs. (8-13) are independent of the potential $U$.
Therefore, an electrically controllable spin-valley valve is realized in AFSL.
The window for a specific spin-valley-polarized state can be effectively switched from on to off and vice versa by adjusting the gate voltages [see Figs. 4(d) and 4(e)].

\begin{figure}
\includegraphics[width=8.0cm,height=5.0cm]{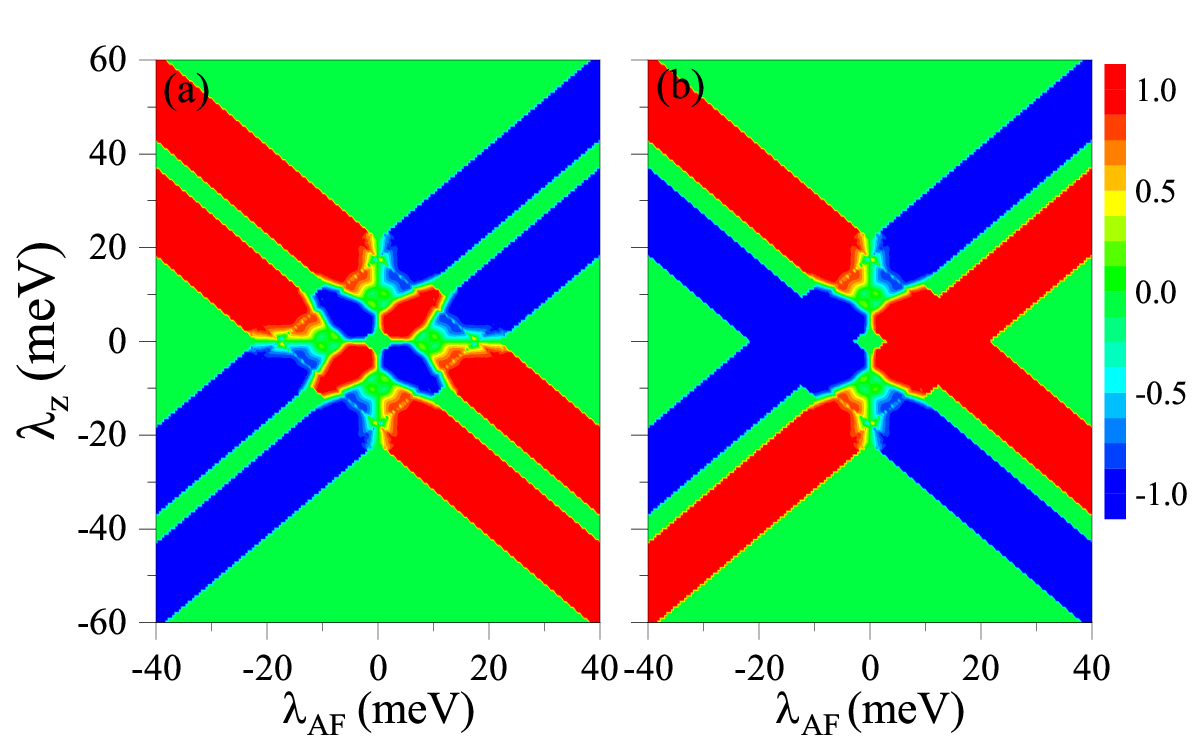}
\caption{ (a) Spin polarization $P_s$ and valley polarization $P_v$ in $(\lambda_{AF}, \lambda_z)$ plane at $U=8meV$. The values of other parameters are the same as these in Fig. 4. }
\end{figure}

\begin{figure}
\includegraphics[width=7.0cm,height=6.0cm]{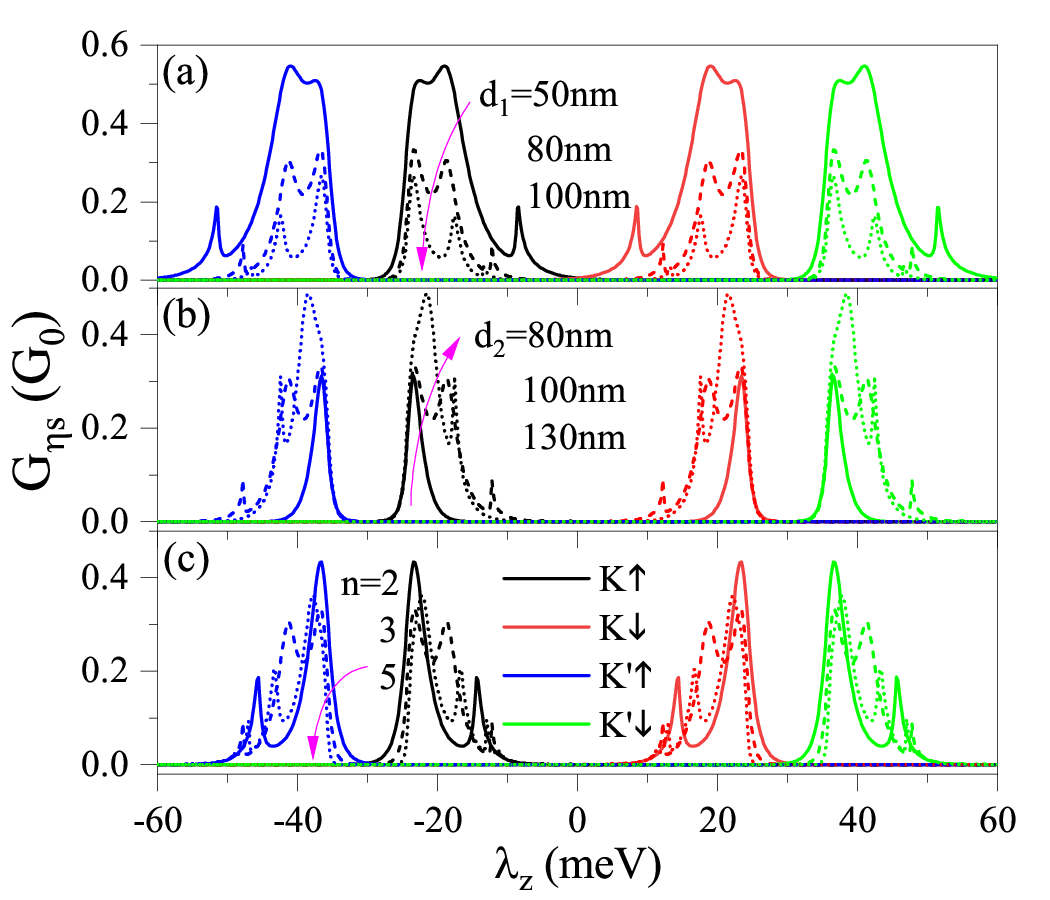}
\caption{ Conductance $G_{\eta s}$ versus electric field $\lambda_z$.
The parameters are set as $n=3$, $\lambda_{AF}=30meV$, $U=8meV$, $d_1=80nm$, and $d_2=100nm$, unless otherwise noted in the figure.  }
\end{figure}

Figs. 5(a) and 5(b) show the spin and valley polarizations $P_{s,v}$ in $(\lambda_{AF}, \lambda_z)$ plane, respectively.
It can be seen that both $P_s$ and $P_v$ have almost ideal platforms in specified areas.
Each platform is supported by a certain spin-valley-polarized state.
There is no conductance in the green area, corresponding to the band gap.
$P_s$ and $P_v$ are antisymmetric with respect to $\lambda_{AF}=0$, consistent with the discussion in Fig. 3.
Dramatically, the symmetry with respect to electric field $\lambda_{z}$ is different from the symmetry with respect to AF exchange field $\lambda_{AF}$, as discussed in Figs. 2-5 and Eqs. (8-13).
By adjusting $\lambda_{AF}$ and $\lambda_z$, both $P_s$ and $P_v$ can be independently switched between $-1$, $0$, and $1$.
The results indicate that the system has effective spin-valley valve effect in a wide range of parameter values.

At the end of this subsection, we discuss the effects of structural parameters $d_1$, $d_2$, and $n$ in Figs. 6(a)-6(c), respectively.
With the increase of barrier width $d_1$ in Fig. 6(a), the conductance is gradually weakened, its region is narrowed, but the resonance becomes remarkable.
On the contrary, as the well width $d_2$ increases in Fig. 6(b), the conductance is enhanced and its region is widened.
As the period number $n$ increases, more conductance peaks appear and the spin-valley polarization becomes more significant, as shown in Fig. 6(c).
The results of Fig. 6 manifest that the structural parameters $d_1$, $d_2$, and $n$ may affect the details of conductance, but do not break the symmetry relationship and the spin-valley valve can well work still.

\subsection{Anisotropic band structure}
Next, we study the anisotropy and spin-valley dependence of energy band for AFSL by solving Eqs. (6) and (7) numerically.
The barrier and well widths are set as $d_1=d_2=100nm$ for convenience in this subsection.

\begin{figure}
\includegraphics[width=8.0cm,height=5.0cm]{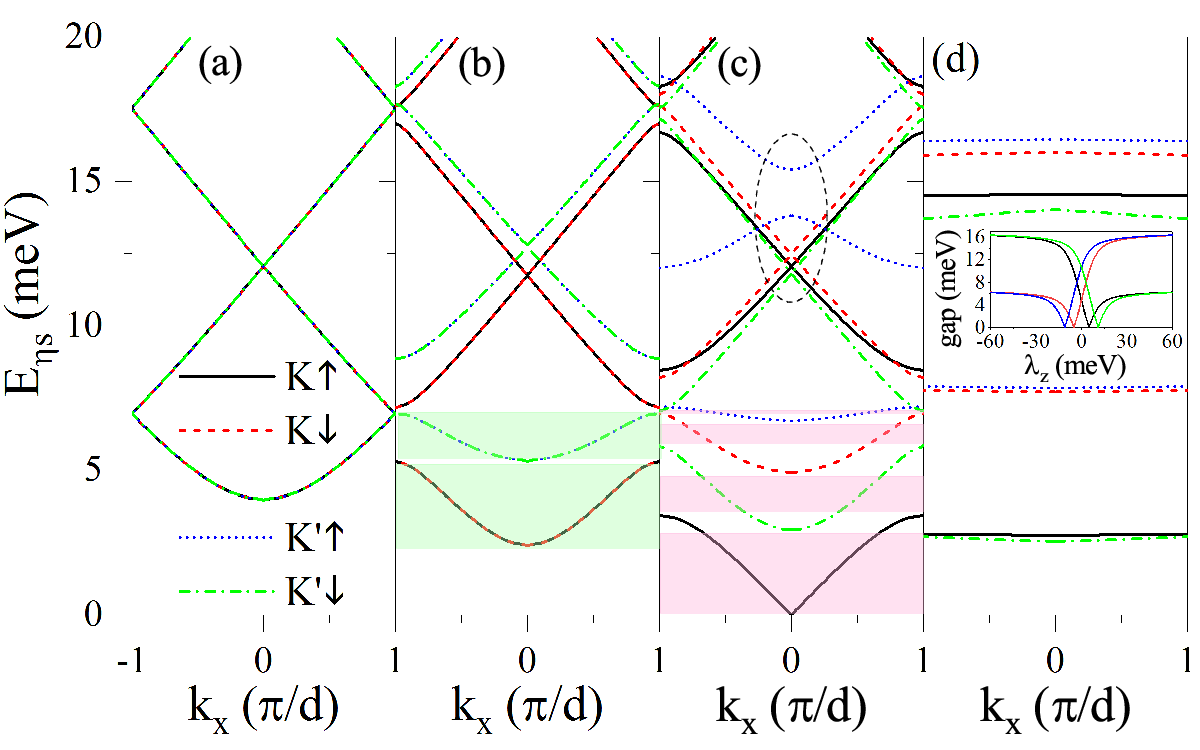}
\caption{ Conduction band versus $k_x$ at (a) $\lambda_z=0$, $\lambda_{AF}=0$;  (b) $\lambda_z=0$, $\lambda_{AF}=3meV$;  (c) $\lambda_z=5meV$, $\lambda_{AF}=3meV$;  (d) $\lambda_z=25meV$, $\lambda_{AF}=3meV$.
The inset of (d) is the band gap versus $\lambda_z$ at $\lambda_{AF}=3meV$.
Other parameters are $\lambda_{SO}=4meV$, $U=0$, and $k_y=0$.}
\end{figure}

Fig. 7 presents the conduction band as a function of $k_x$ when $U=0$ and $k_y=0$. The conduction band and valence band are symmetric about the Fermi level at $U=0$. As expected, at $\lambda_z=\lambda_{AF}=0$ in Fig. 7(a), the bands for the four kinds of electronic states are completely degenerate with a band gap of $2\lambda_{SO}$.
The appearance of $\lambda_{AF}$ would make the minibands of $K$ valley shift down, but the minibands of $K'$ valley shift up [see Fig. 7(b)], because of the local band gap $\Delta_{\eta}=\lambda_{AF} - \eta \lambda_{SO}$ in AF region.
This enables the observation of perfectly valley-polarized transport in AFSL system, as discussed in subsection III A.
Note that such a phenomenon does not exist in ferromagnetic SL, which would cause the minibands with different spins to move in the opposite direction of energy \cite{Missault2}.
Due to SOC, the combined effect of $\lambda_{AF}$ and $\lambda_z$ also eliminates spin degeneracy, and so the minibands of different spin-valley electrons appear in different energy ranges, as shown in Fig. 7(c).
For the proper energy, only one miniband for a given spin-valley state could be observed, and this leads to the spin-valley valve effect in subsection III B.
When $\lambda_z+s\lambda_{AF}=2 \eta s \lambda_{SO}$, such as $\lambda_z=5meV$ and $\lambda_{AF}=3meV$, the band gap of the $K\uparrow$ ($\eta=s=1$) electron is closed and the Dirac point is formed at $E=k_x=k_y=0$, while other spin-valley electrons still have a band gap.
It also opens up the finite-energy crossed points near $K'\uparrow$ valley [see the ellipse in Fig. 7(c)].
At the same time, the first conduction band of $K'\uparrow$ becomes almost flattened.
Remarkably, with the further increase of $\lambda_{AF}$ or $\lambda_z$, all the electrons can get flattened minibands at different positions in the low energy range [see Fig. 7(d)], which means that the corresponding group velocity along the $x$ direction tends to zero.
The flattened minibands arise from the bound states formed inside the potential wells of AFSL.
The bound states decay exponentially in the AF barriers. The negligible coupling between the bound states in the adjacent SL periods thus manifests as flattened minibands in the band structure \cite{MJung}.
The inset of Fig. 7(d) exhibits the band gap as a function of $\lambda_z$.
One may find that by adjusting $\lambda_z$ the band gap of a particular spin-valley electron can be closed.
As $\lambda_z$ increases, each band gap first linearly increases and then gradually tends to saturation, which corresponds to the flattened miniband.

\begin{figure}
\includegraphics[width=8.0cm,height=5.0cm]{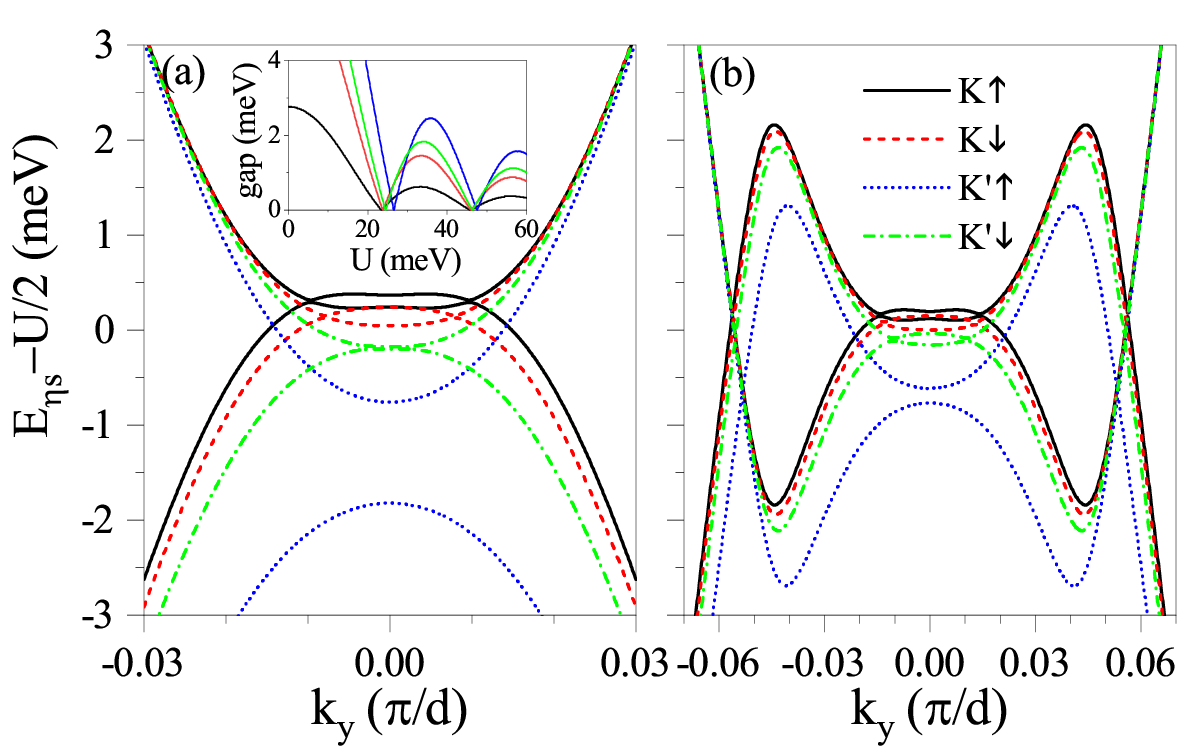}
\caption{ Energy band versus $k_y$ at (a) $U=24.5meV$ and (b) $U=47meV$.
The inset of (a) is the band gaps at original Dirac point versus $U$.
Other parameters are $k_x=0$, $\lambda_z=2meV$, $\lambda_{AF}=3meV$, and $\lambda_{SO}=4meV$.}
\end{figure}

It has demonstrated that the extra zero-energy Dirac points would appear in graphene SL \cite{Brey, Barbier}.
Moreover, the AF exchange field and electric field can destroy the spin and valley degeneracies.
Therefore, it is necessary to study the dependence of extra Dirac points on spin and valley degrees of freedom in AFSL.
Fig. 8 presents the energy band as a function of $k_y$ when (a) $U=24.5meV$ and (b) $U=47meV$.
$U=24.5meV$ is the critical value of $K'\downarrow$ electron that generates a new Dirac point.
From Fig. 8(a) we can see that $K'\downarrow$ electron has one Dirac point at $k_y=0$, both $K\uparrow$ and $K\downarrow$ electrons have two Dirac points at $k_y\neq 0$, but $K'\uparrow$ electron has no Dirac point.
These Dirac points occur in different positions and strongly depend on the spin and valley indexes.
Note that the Dirac cone for $K'\downarrow$ electron tends to flattened in the $k_y$ direction, suggesting that the velocity along the $y$ direction is normalized to zero.
In the condition of $d_1=d_2$ and $k_x=0$, we may obtain the spin-valley-dependent location ($E_0$, $k_{y0}$) of Dirac point by solving Eqs. (6) and (7), that is
\begin{align}
&E_0=\frac{U}{2} + \frac{\lambda_{SO}^2-(\lambda_z + s \lambda_{AF} - \eta s \lambda_{SO})^2}{2U}, \\
&k_{y0}=\pm \sqrt{\frac{E_0^2-\lambda_{SO}^2}{(\hbar v_F)^2}-\frac{(2m\pi)^2}{d^2}},
\end{align}
with the positive integer $m$.
The location ($E_0$, $k_{y0}$) is effectively controlled by the AF exchange field $\lambda_{AF}$ and the electric field $\lambda_z$.
The inset of Fig. 8(a) shows the variation of band gap at $k_y=0$ as a function of $U$.
When a certain gap is closed, a new Dirac point will be created at $k_y=0$, such as $K'\downarrow$ electron at $U=24.5meV$ [see the green curve in Fig. 8(a)].
As $U$ increases, the new Dirac point splits into a pair and they move in opposite directions away from the point $k_y=0$ but always keeping $k_x=0$.
The Dirac points located at ($E_0$, $k_{y0}$) are symmetrically distributed with respect to $k_y=0$, which can be understood by the operator $\mathcal{M}_1$.
Obviously, the critical value of the new Dirac point is different for different spin-valley electrons.
With the further increase of $U$, the gap will reopen and close again, suggesting that more extra Dirac points will appear [see Fig. 8(b)].

\begin{figure}
\includegraphics[width=8.0cm,height=6.0cm]{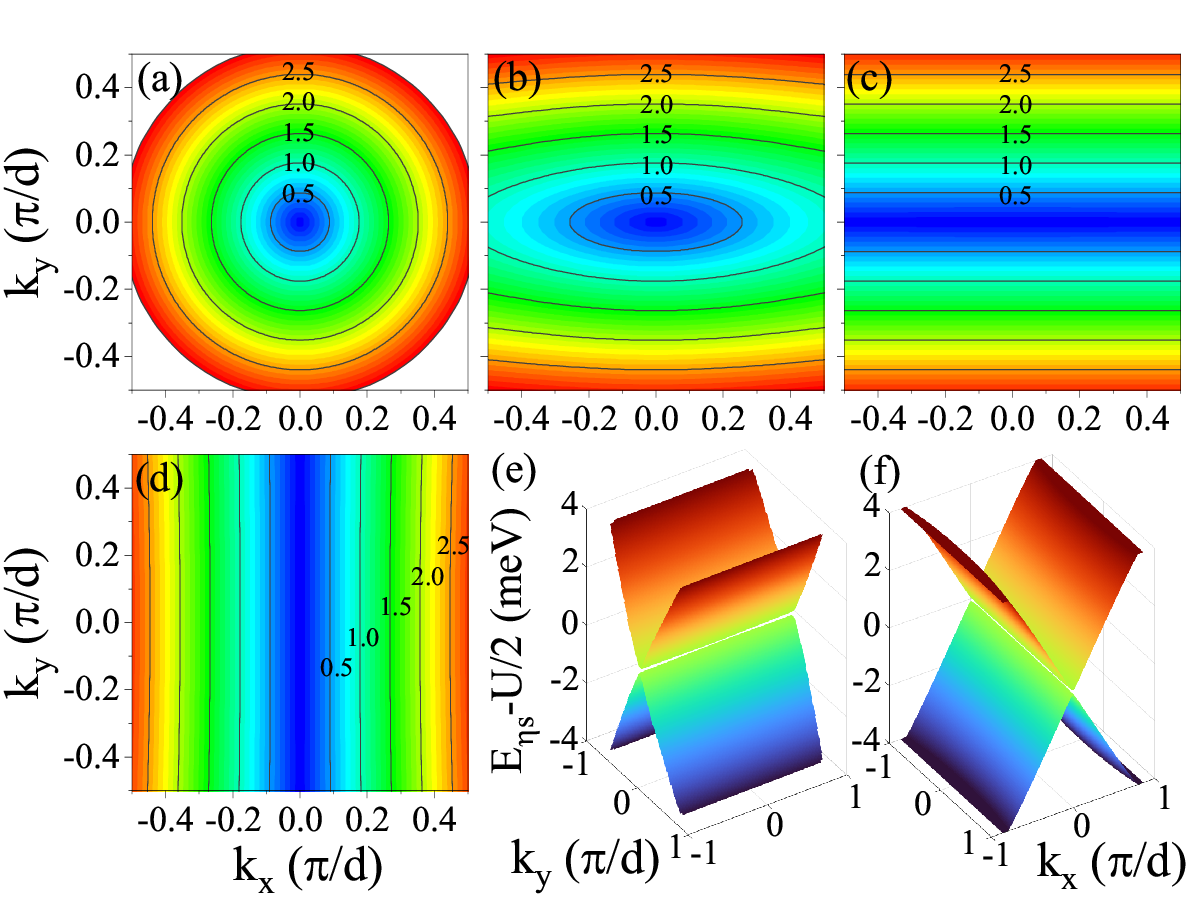}
\caption{ (a)-(f) Contour plots of the first conduction miniband for $K\uparrow$ electron at (a) $U=0$, $\lambda_{SO}=0$; (b) $U=0$, $\lambda_{SO}=10meV$; (c) $U=0$, $\lambda_{SO}=25meV$; (d) $U=47meV$, $\lambda_{SO}=4meV$.
(e) and (f) are the 3D plots of the first conduction and valence minibands for (c) and (d), respectively.
The values of $\lambda_z$ and $\lambda_{AF}$ satisfy $\lambda_z + \lambda_{AF} = 2 \lambda_{SO}$.
The energy unit is $meV$ in (a)-(d). }
\end{figure}

Figs. 9-11 discuss an anomalous anisotropy of band structure caused by SOC and AF exchange field in AFSL, which is different from the anisotropic behaviour in graphene SL \cite{Park2, YLi}.
Results in Fig. 7 have proved that the Dirac point is formed when $\lambda_z+s\lambda_{AF}=2 \eta s \lambda_{SO}$ and $U=0$.
In the following, let us take $K\uparrow$ electron as an example and discuss the anisotropy of Dirac point when $\lambda_z+\lambda_{AF}=2 \lambda_{SO}$.
Figs. 9(a)-9(c) show the contour plots of the first conduction miniband at $U=0$ for different values of $\lambda_{SO}$.
For silicene, germanene, and stanene, the SOC strength $\lambda_{SO}$ is about $4meV$, $43meV$, and $100meV$, respectively \cite{CCLiu, YXu}, and it can be regulated via chemical functionalization \cite{YXu}.
When $\lambda_{SO}=0$, the system is simplified to pristine graphene and the band is expected to be isotropic in the Brillouin zone, as shown in Fig. 9(a).
The group velocity of states near the Dirac point is parallel to wavevector $\bm{k}$ and has a constant value $v_F$.
However, as SOC $\lambda_{SO}$ appears, the band becomes anisotropic [see Fig. 9(b)].
Comparing Figs. 9(a) and 9(b), the change of isopotential contours indicates that the group velocity is anisotropically renormalized and it strongly depends on the direction of $\bm{k}$.
In particular, the velocity perpendicular to the periodic direction of AFSL $v_y$ is not renormalized at all, that is $v_y=v_F$, but the velocity parallel to the periodic direction $v_x$ is decreased.
Importantly, such an anisotropy is contrary to that in graphene SL, where $v_y$ is reduced, whereas $v_x$ keeps constant \cite{Park2, YLi}.
The anisotropy would become more significant with the increase of $\lambda_{SO}$.
Taking $\lambda_{SO}=25meV$ for instance in Fig. 9(c), there is hardly any dispersion along the $k_x$ direction, while the dispersion along the $k_y$ direction is always the same as that at $\lambda_{SO}=0$.
The components $v_{x,y}$ of velocity can be defined as
\begin{align}
v_x / v_F = \partial E_{\eta s} / \partial k_x ,  \quad  v_y / v_F = \partial E_{\eta s} / \partial k_y.
\end{align}
The velocities are $v_x \approx 0.013v_F$ and $v_y \approx v_F$ at $\lambda_{SO}=25meV$.
The velocity along the $x$ direction $v_x$ almost vanishes.
Fig. 9(e) gives the 3D plot of the first conduction and valence minibands for Fig. 9(c).
We can clearly see an interesting Dirac line along the $k_x$ direction instead of a Dirac point.
The band possesses a highly flattened energy dispersion in the $k_x$ direction.

The appearance of potential $U$ will change the anisotropy property.
Figs. 9(d) and 9(f) display the contour plot and 3D plot of the miniband when $U=47meV$ and $\lambda_{SO}=4meV$.
In this situation, even though $\lambda_{SO}$ is finite, the energy dispersion disappears along $k_y$ direction but it has no change along $k_x$ direction, in agreement with the ones in graphene SL \cite{Park2, YLi}.
As a result, the anisotropic velocities at Dirac point become $v_x \approx v_F$ and $v_y \approx 0.017v_F$.
Distinctly, there are two different anisotropic behaviours in Figs. 9(c,e) and 9(d,f).
By adjusting $U$ and $\lambda_{SO}$, we can achieve an electron supercollimation along the $x$ or $y$ directions.

\begin{figure}
\includegraphics[width=8.0cm,height=3.3cm]{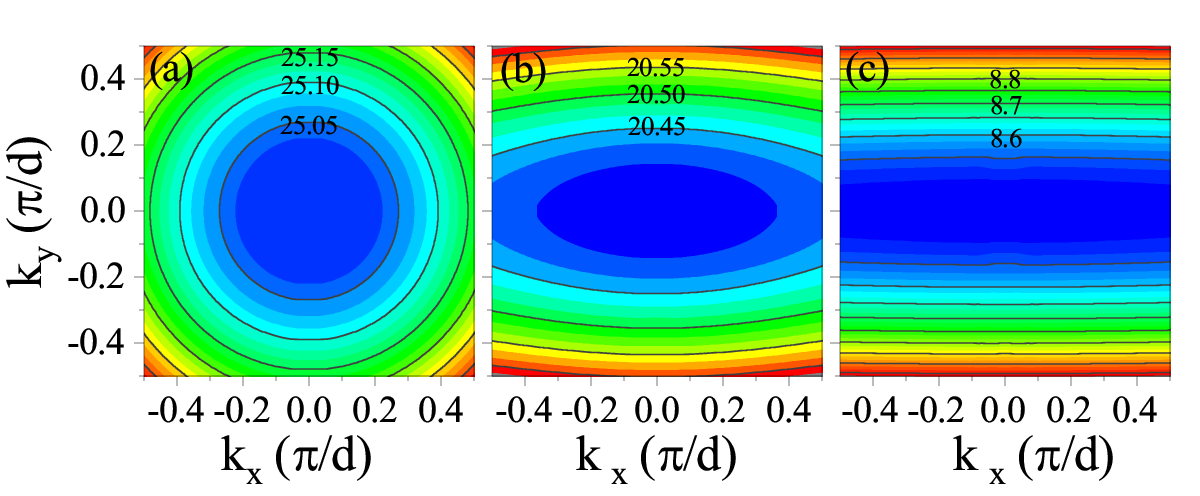}
\caption{ Contour plots of the first conduction miniband for $K\uparrow$ electron when $\lambda_z + \lambda_{AF} = $ (a)  $0$, (b) $6meV$, and (c) $20meV$. The energy unit is $meV$.
Other parameters are $U=0$ and $\lambda_{SO}=25meV$. }
\end{figure}

In order to discuss the reason of the former anisotropy at Dirac point, Fig. 10 shows the contour plots of the first conduction miniband before the formation of Dirac point at $\lambda_{SO}=25meV$ for different values of $\lambda_z + \lambda_{AF}$.
We can see that when $\lambda_z + \lambda_{AF}=0$ in Fig. 10(a), the band is clearly isotropic.
The group velocity at the bottom of conduction band is almost zero due to the bound states inside the potential wells of AFSL.
However, with the appearance of $\lambda_z + \lambda_{AF}$, the band becomes anisotropic and moves downward along the energy.
The contours in the $k_y$ direction become more and more dense as $\lambda_z + \lambda_{AF}$ increases [see Figs. 10(b), 10(c), and 9(c) with $\lambda_z + \lambda_{AF}=6$, $20$, and $50meV$, respectively].
The profile of energy dispersion $E_{\eta s} (k_y)$ along $k_y$ direction gradually changes from a parabolic shape to a linear shape, which means that the velocity $v_y$ gradually increases to $v_F$.
Conversely, the contours in the $k_x$ direction become thinner and the band $E_{\eta s} (k_x)$ becomes flattened, meaning that $v_x$ becomes smaller and tends to zero.
Therefore, the anisotropic band with renormalized velocity $v_x$ is formed due to the bound states induced by SOC, $\lambda_{AF}$, and $\lambda_z$.
Whereas, the anisotropic band with renormalized velocity $v_y$ results from the chiral nature of the states \cite{Park2, YLi}.

\begin{figure}
\includegraphics[width=8.0cm,height=5.0cm]{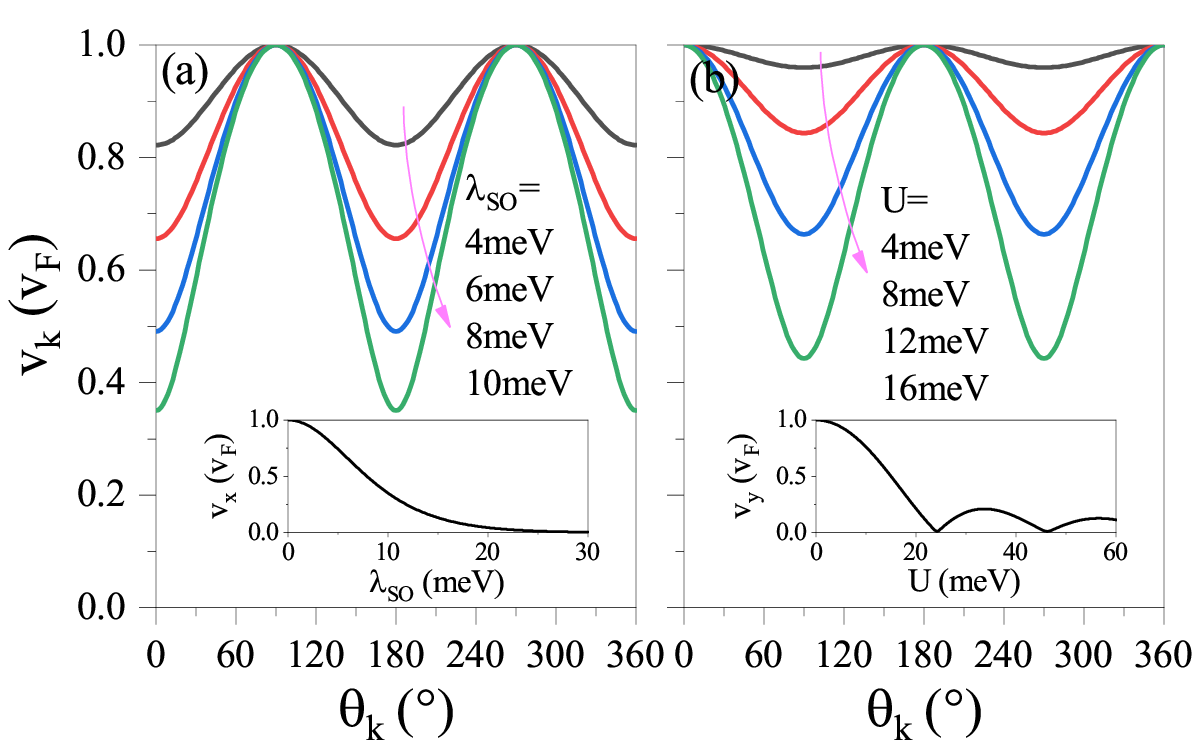}
\caption{ Anisotropic velocity $v_k$ versus the angle $\theta_k$ at (a) $U=0$ and (b) $\lambda_{SO}=4meV$ for $K\uparrow$ electron.
The inset of (a) is velocity component $v_x$ versus $\lambda_{SO}$ at $U=0$.
The inset of (b) is velocity component $v_y$ versus $U$ at $\lambda_{SO}=4meV$.
The values of $\lambda_z$ and $\lambda_{AF}$ satisfy $\lambda_z + \lambda_{AF} = 2 \lambda_{SO}$.}
\end{figure}

We may study the variation of two kinds of anisotropic velocities with the angle.
The velocity $v_k(\theta_k)$ can be obtained from $v_x$ and $v_y$, namely,
\begin{align}
v_k (\theta_k) = v_x \cos \theta_k + v_y \sin \theta_k,
\end{align}
where $\theta_k$ is the angle of wavevector $\bm{k}$ from the periodic direction $x$.
Figs. 11(a) and 11(b) exhibit the angle dependence of the velocity $v_k(\theta_k)$ caused by the SOC and the chirality, respectively.
For the SOC induced anisotropy in Fig. 11(a), the velocity $v_k$ reaches its maximum value equal to $v_F$ at $\theta_k=90^\circ$ and $270^\circ$, i.e., $v_y$, and it has minimum value at $\theta_k=0^\circ$ and $180^\circ$, i.e., $v_x$.
As $\lambda_{SO}$ increases, $v_x$ decreases gradually, while $v_y$ remains constant regardless of the SOC strength.
The inset of Fig. 11(a) shows that $v_x$ decreases monotonically with $\lambda_{SO}$.
When $\lambda_{SO}>25meV$, one may get $v_x \approx 0$ and $v_y \approx v_F$, realizing a good supercollimation effect in the $y$ direction.
On the contrary, for the chirality induced anisotropy in Fig. 11(b), $v_y$ is decreased but $v_x$ keeps unchanged as $U$ increases.
The inset of Fig. 11(b) indicates that $v_y$ oscillates damply with $U$. 
At the critical value for the generation of new Dirac point, we have $v_y=0$ and create a supercollimation in the $x$ direction.

It should be noted that when SOC and potential $U$ exist simultaneously in Figs. 9(d), 9(f), and 11(b), the chiral nature of the states plays a leading role in the formation of anisotropic band.
The direction of the renormalized group velocity can be regulated depending on SOC and $U$.
In addition, as discussed in Figs. 9-11, when $\lambda_z + \lambda_{AF} = 2 \lambda_{SO}$, $K\uparrow$ electron realizes the anisotropy and the supercollimation, whereas other spin-valley electrons do not have these properties. 
When $\lambda_z + s \lambda_{AF} = 2 \eta s \lambda_{SO}$, a certain spin-valley electron can achieve the anisotropy. 
Therefore, the anisotropy is spin-valley dependent due to SOC and AF exchange fields.

Finally, we discuss the feasibility of the AFSL with graphene or other 2D hexagonal lattice as a platform in experiment. 
The proposed model should be general and applicable to systems in which antiferromagnetism is either intrinsic to the materials \cite{JJung, Lado} or induced by proximity \cite{XLi, Hogl, Ezawa2, PWei, Gargiani}.
Density functional theory predicts that AF hexagonal lattice can be naturally realized by depositing the hexagonal lattice on monolayer MnPX$_3$ (X = S, Se) \cite{XLi, Hogl}. Because bulk MnPX$_3$ is a layered compound, and only the top monolayer is important for proximity effects, the AF order with strength on $meV$ scale can be experimentally induced using a MnPX$_3$ film \cite{Hogl}.
Alternatively, the AF order can be also realized by sandwiching the buckled hexagonal lattice between two different ferromagnets \cite{Ezawa2}, such as EuS \cite{PWei}.
It has been experimentally demonstrated that the ultra-thin Fe/graphene/Co films grown on Ir(111) exhibit a robust perpendicular AF exchange field which is stable above room temperature \cite{Gargiani}. 
The AF exchange field can be controlled by adjusting the magnetization direction, geometric configuration, and the distance between the AF substrates \cite{XLi, Hogl, PWei, Gargiani}.
Furthermore, the high quality superlattice structures based on 2D materials have been experimentally realized within the existing technology \cite{YLi, TLi, Ghorashi}.

In addition, due to the imperfection of experiment, the strengths of AF exchange fields on the two sublattices may be unequal, which are set as $\lambda_{AF1}$ and $\lambda_{AF2}$. Then the system's Hamiltonian can be rewritten as 
\begin{align}
H_{\eta s}=H_0 + (\lambda_z - \eta s \lambda_{SO}) \sigma_z + s \lambda_{AF1} \sigma_+ + s \lambda_{AF2} \sigma_- + U,
\end{align}
where $\sigma_\pm = (\sigma_z \pm \sigma_0)/2$ and $\sigma_0$ is unit matrix.
The eigenvalue becomes $E_{\eta s} = U_{eff}  \pm \sqrt {\Delta_{\eta s}'^2 + (\hbar v_F k_F)^2}$ with $U_{eff}=U + s (\lambda_{AF1} - \lambda_{AF2})/2$ and $\Delta_{\eta s}'=\lambda_z - \eta s \lambda_{SO} + s (\lambda_{AF1} + \lambda_{AF2})/2$.
When $\lambda_{AF1} \neq \lambda_{AF2}$, the electrons will experience a spin-dependent effective potential $U_{eff}$ and so the spin-dependent conductances will be decreased or increased. 
The condition of Dirac point will become $\lambda_z + s(\lambda_{AF1}+\lambda_{AF2})/2=2\eta s \lambda_{SO}$, instead of $\lambda_z + s \lambda_{AF}=2\eta s \lambda_{SO}$.
However, the anisotropic behaviour and spin-valley valve always exist.
Therefore, the theoretical results should be available in experiment.

\section{Conclusion}
In summary, we proposed a symmetry-protected spin-valley valve effect and found an anomalous anisotropy in AFSL.
The property of perfect spin-valley-polarized minibands and conductance is discussed based on the system's symmetry.
The obtained spin-valley polarization and symmetry relations should be beneficial to the experimental design of AF device.

Furthermore, the SOC gives rise to an anisotropic Dirac cone in infinite AFSL, where the group velocity along the $x$ direction is greatly renormalized and  monotonically suppressed as a function of SOC, opposite to the anisotropy caused by the chirality in graphene SL.
Remarkably, the strength and the direction of anisotropy in energy dispersions can be controlled by changing SOC and the potential. 
The anisotropy, supercollimation, and extra Dirac points strongly depend on the spin and valley indexes due to SOC.
These results demonstrate that AFSL can be used as an alternative approach to engineer the anisotropic 2D materials.

\section*{Acknowledgments}
This work was supported by the National Natural Science Foundation of China (Grants No. 11974153, No. 11921005, and No. 12374034), the Innovation Program for Quantum Science and Technology (Grant No. 2021ZD0302403), and the Strategic Priority Research Program of Chinese Academy of Sciences (Grant No. XDB28000000).

\end{CJK*}
\end{document}